\documentclass[floats,prb,aps,twocolumn]{revtex4}
\usepackage{amsmath}
\usepackage{graphicx}

\newcommand{\slashi}{{i\kern-0.3em/}}
\newcommand{\slashone}{{1\kern-0.4em/}}

\begin{document}
\title{Physical mechanism for a kinetic energy driven zero-bias anomaly in the Anderson-Hubbard model} 
\author{R. Wortis} \email{rwortis@trentu.ca}
\author{W. A. Atkinson} 
\affiliation{Trent University, 1600 West Bank Dr., Peterborough ON,
  K9J 7B8, Canada} 
\date{\today}
\begin{abstract}
The combined effects of strong disorder, strong correlations and hopping in the Anderson-Hubbard model have been shown to produce a zero bias anomaly which has an energy scale proportional to the hopping and minimal dependence on interaction strength, disorder strength and doping.
Disorder-induced suppression of the density of states for a purely local interaction is inconsistent with both the Efros-Shklovskii Coulomb gap and the Altshuler-Aronov anomaly, and moreover the energy scale of this anomaly is inconsistent with the standard energy scales of both weak and strong coupling pictures.
We demonstrate that a density of states anomaly with similar features arises in an ensemble of two-site systems, and we argue that the energy scale $t$ emerges in strongly correlated systems with disorder due to the mixing of lower and upper Hubbard orbitals on neighboring sites.  
\end{abstract}
\maketitle

{\em Introduction.}
Since the discovery of high temperature superconductivity, there has been widespread interest in the physics of doped Mott insulators.  These materials, mainly transition metal oxides, have electronic properties which are dominated by electron-electron correlations and which can be tuned by chemical doping.  
Recently, there have been indications that the disorder introduced by doping can have important effects on magnetic\cite{Alloul2009}, superconducting\cite{Fischer2007}, and transport properties\cite{Rullier2008}, especially near the Mott metal-insulator transition.
In particular,  experiments have found disorder-induced suppression 
of the density of states (DOS) at the Fermi energy $\varepsilon_F$ 
in several transition metal oxides\cite{Sarma1998,Naqib2005,Ray2006,Maiti2007}.  
Interpretation of these results is generally given in terms of Altshuler-Aronov (AA) zero bias anomalies (ZBA), which are known to occur in disordered metals\cite{Altshuler1985}, as well as the Efros-Shklovskii (ES) Coulomb gap, which arises in the atomic limit.\cite{Efros1975a}
However, while low disorder samples often give results consistent with AA, the pattern in strongly disordered samples is a failure to fit either of these pictures.  This highlights the importance of developing a new picture which captures the physics of strong correlations and strong disorder.

In this work we address the physics of the ZBA in a widely used model for strongly correlated systems with disorder, the Anderson-Hubbard model (AHM).  The Hamiltonian is
\begin{equation}
\hat H = \sum_{i,j,\sigma} t_{ij} \hat c^\dagger_{i\sigma} \hat c_{j\sigma}
+ \sum_i \left (
\epsilon_i \hat n_i + U \hat n_{i\uparrow}\hat n_{i\downarrow}
\right ),
\label{eq:Ham}
\end{equation}
where $t_{ij}=-t$ for nearest-neighbor sites $i$ and $j$, and is zero otherwise, and $\hat c_{i\sigma}$ and $\hat n_{i\sigma}$ are the annihilation and number operators for lattice site $i$ and spin $\sigma$, and $\epsilon_i$ is the energy of the orbital at site $i$.
Disorder is introduced by choosing $\epsilon_i$ from a  uniform distribution $\epsilon_i \in [-\frac{\Delta}{2},\frac{\Delta}{2}]$.

The AHM has been extensively studied by numerous methods, including Hartree-Fock\cite{Heidarian2003,Fazileh2006}, dynamical mean field theory\cite{Miranda2005,Semmler2010,Song}, and exact methods\cite{Kotlyar2001,Srinivasan2003}.  
However, the existence of a suppression of the DOS at the Fermi level (a ZBA) in low dimensions was only recently 
established conclusively via exact diagonalization and quantum Monte Carlo calculations\cite{Chiesa2008,Shinaoka2009}.
%In Ref.~\onlinecite{Chiesa2008} the ZBA appears as a V-shaped suppression of the DOS at $\varepsilon_F$.
Ref.~\onlinecite{Chiesa2008} found a V-shaped suppression of the DOS at $\varepsilon_F$.
A narrower soft gap, of width $\sim 0.1t$, was also found by exact diagonalization of the one-dimensional AHM,\cite{softgap2} and was attributed to long-range correlations.\cite{Shinaoka2009}  
However the V-shaped anomaly in Ref.~\onlinecite{Chiesa2008} with an energy scale $\sim t$ remains unexplained.
This anomaly is distinct from the Mott gap, as demonstrated at half filling by its stability at increasing disorder strengths after the Mott gap has filled in, and by its persistence away from half filling.

These properties raise a number of questions.
First, the suppression of the DOS at the Fermi level by a purely local interaction is inconsistent with both the ES and AA pictures.  
The ES Coulomb gap comes specifically from the long range $1/r$ interaction, whereas the purely local interaction in the AHM produces a flat DOS in the atomic limit.
The AA anomaly, at the mean-field level, comes from exchange terms which vanish for the AHM.
Second, the energy scale of the AHM anomaly is of order $t$, in contrast both to the Hartree-Fock energy scale $U$ and to the strong coupling scale $J \propto t^2/U$.
An alternative to the existing frameworks is required to gain insight into these behaviors.

In this paper we present a simple physical mechanism which results in a ZBA with energy scale $t$ in the AHM: the mixing of the lower Hubbard orbital on one site with the upper Hubbard orbital on a neighboring site, an effect unique to strongly correlated systems.
We demonstrate that the DOS of an ensemble of two-site systems displays a ZBA which shares key features with that found in the lattice.\cite{Chiesa2008}
By focusing on the subset of the ensemble which contributes most strongly to the anomaly for $\Delta > U$ near half filling, we isolate the underlying physics.  
The mechanism in this parameter range is confirmed by an analytic calculation of the DOS to leading order in $t/U$.
We comment on the behavior outside this parameter range and on the implications for experiments.

{\em Ensemble of molecules}
In general, disorder reduces the importance of kinetic energy relative to interactions, and in the limit of strong disorder 
hopping causes only minimal changes to the atomic spectrum.
Therefore, an obvious starting point for a strongly disordered system is the atomic limit.
Given that the atomic limit in the case of a purely local interaction has no ZBA, 
a natural next step is to consider an ensemble of two-site systems, referred to here as molecules.
In considering molecules containing just two sites, we can expect to capture only the highest energy scale of the ZBA but not the low energy behavior coming from longer range correlations.  We therefore focus our attention on the width rather than the depth of the resulting anomaly.

\begin{figure}
\begin{center}
%\begin{tabular}{c}
%\includegraphics[width=\columnwidth]{v4_agr_to_eps_to_pdf}
%\\
%\includegraphics[width=\columnwidth]{v7_agr_to_eps_to_pdf}
%\end{tabular}
\includegraphics[width=\columnwidth]{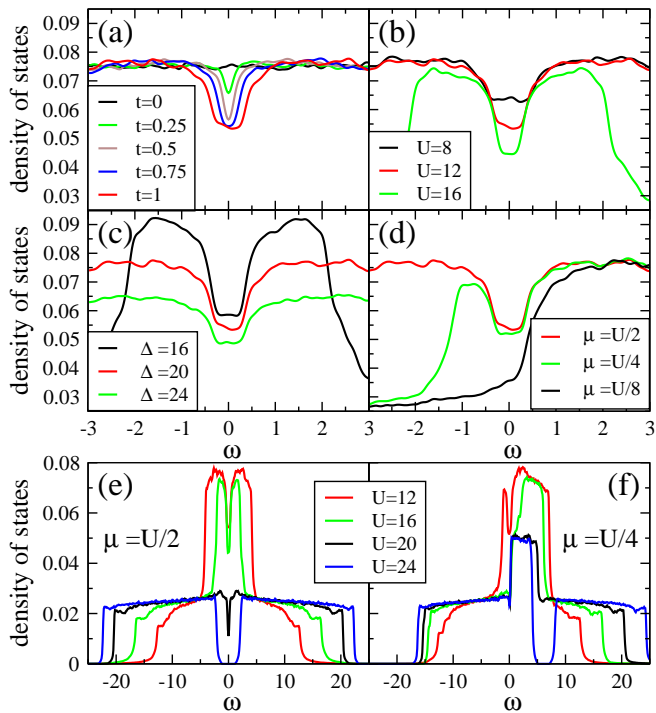}
\caption{(color online) The DOS of the AHM for an ensemble of 500,000 two-site systems.
$t=1$, $U=12$, $\Delta=20$, and $\mu=U/2$ unless otherwise specified. 
(a)-(d) show the variation of the ZBA with (a) $t$, (b) $U$, (c) $\Delta$, and (d) $\mu$.
Displaying the full frequency range, (e) and (f) show that at half filling the ZBA goes away and is replaced by the Mott gap for $U>\Delta$,
and below half filling the ZBA gives way to a step in the DOS.
This step edge is also seen at the lowest doping in (d).
} 
\label{fig:ensemble_dos}
\end{center}
\end{figure}

The molecular Hamiltonian is Eqn. (1) with $i,j=1,2$ only and site energies chosen from a uniform distribution.
The ensemble-averaged DOS is smooth and, at half filling, preserves electron-hole symmetry.
Fig. 1 shows the resulting DOS for a range of parameters, with panels (a)-(d) focusing near the Fermi level.
Panel (a) shows a ZBA opening with the addition of hopping.
Panels (b) and (c) show significant variation in the depth of the anomaly with interaction strength $U$ and disorder strength $\Delta$, but essentially no variation of the width.  
In panel (d) the anomaly is the same at half-filling ($\mu=U/2$) and below ($\mu=U/4$). 
%Panels (e) and (f) show the full frequency range and demonstrate (e) the opening of the Mott gap for $U>\Delta$ at half filling and (f) the shift in the placement of the Fermi level from the central plateau to a step edge for large $U$ values below half filling.  
%This step edge is also seen at the lowest doping in (d).
%Panel (e) emphasizes that the ZBA is distinct from the Mott gap.
While the shape of the anomaly is different from that seen in the lattice,\cite{Chiesa2008} the {\em width} of the anomaly shares many features with that of the lattice:  Most notably, the width is not only linear in $t$ but roughly equal to $t$.  The ZBA is also independent of $U$, $\Delta$, and $\mu$.  The advantage here is that the system is sufficiently simple that we can explore the anomaly's origin.

\begin{figure}
\begin{center}
%\begin{tabular}{cc}
%\includegraphics[width=1.5in]{temp_02}
%&
%\includegraphics[width=1.5 in]{site_energies_b_v3}
%\end{tabular}
\includegraphics[width=\columnwidth]{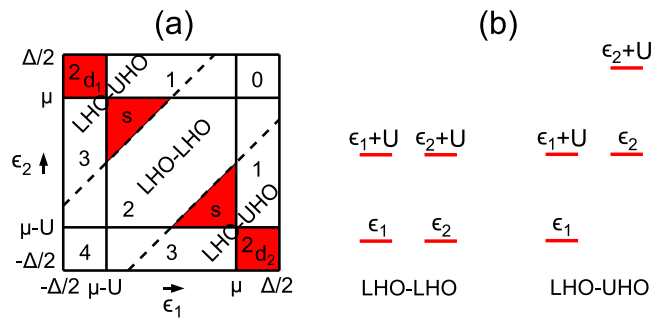}
\caption{(color online) (a) The phase space of a molecule with regions labeled according to the number of particles in the atomic ground state and according to the alignment of the atomic orbitals as shown in (b).
%(c) \& (d) Examples of this phase space, emphasizing $|d\rangle$ and $|s\rangle$ molecules, for $U=12$ and $\Delta=20$, and $\mu=U/2$ and $\mu=U/4$ respectively.
}
\label{fig:site_energies}
\end{center}
\end{figure}

{\em Origin of the anomaly}
The addition of hopping changes the atomic DOS of each molecule, and an exhaustive description of all these changes is not sufficiently simple to be useful.
Here we focus on those molecules which make the dominant contribution to the anomaly for $\Delta>U$ near half filling.
We begin by describing two useful criteria for grouping the molecules, and then explore how an anomaly develops from one of the resulting groups by a mechanism which is unique to strongly correlated systems.

Recalling that the poles in the Green's function occur at the transition energies between many-body states and that the allowed transitions depend on the number of particles in the ground state,
a first useful grouping of the molecules is in terms of the number of particles in the atomic ground state.
Fig.~\ref{fig:site_energies}(a) shows all possible values of the site potentials $\epsilon_1$ and $\epsilon_2$ divided into regions according to the number of particles in the atomic ground state.
The changes caused by $t$ will be greatest when two atomic orbitals are close in energy, and the grouping of molecules may be further refined by distinguishing two ways in which this alignment may occur.  In the atomic limit, interactions split the resonance at each site into a lower Hubbard orbital (LHO) with energy $\epsilon_i$ and an upper Hubbard orbital (UHO) with energy $\epsilon_i+U$.  
In a molecule, the LHO of one site may be more closely aligned with the LHO of the other site (LHO-LHO in Fig.~\ref{fig:site_energies}(b)) or it may be more closely aligned with the UHO of the other site (LHO-UHO in Fig.~\ref{fig:site_energies}(b)).\cite{phase_space_range} 
When $\Delta>U$ near half filling, the largest contribution to the anomaly comes from molecules with 2-particle atomic ground states and LHO-UHO alignment.  Fig.~\ref{fig:site_energies}(a) marks these molecules in red, and labels those in which site $i$ is doubly occupied in the atomic limit $|d_i\rangle$ and those in which both sites are singly occupied $|s\rangle$.

\begin{figure}
\begin{center}
\includegraphics[width=\columnwidth]{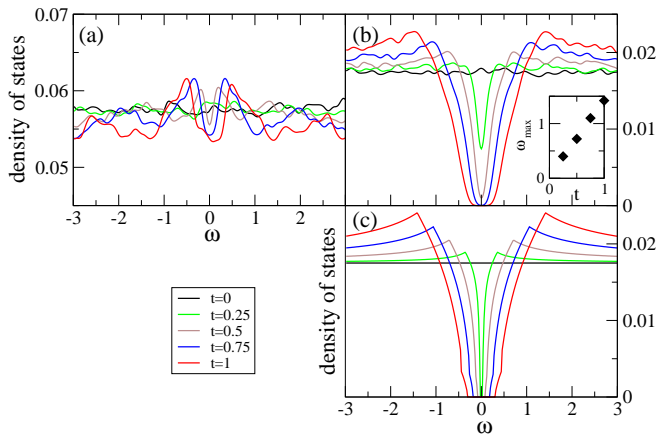}
\caption{(color online) Contribution to the DOS (a) from all molecules {\em except} $|d\rangle$ and $|s\rangle$ and (b) from $|d\rangle$ and $|s\rangle$ molecules.%\cite{hatch}  
Inset shows position of maximum in DOS vs. $t$.  (c) 
Analytic expression (Eqn.~(\ref{analyticeqn})) for $|d\rangle$ and $|s\rangle$ contribution.
% All results are for $U=12$, $\Delta=20$ and $\mu=U/2$.
} 
\label{fig:2ptcle_contrib}
\end{center}
\end{figure}

Fig.~\ref{fig:2ptcle_contrib}(a) shows the contribution to the DOS of all molecules in the ensemble {\em except} the $|d\rangle$ and $|s\rangle$ molecules.
While there is a small ZBA, it is dwarfed by that of 
the $|d\rangle$ and $|s\rangle$ molecules, shown in Fig.~\ref{fig:2ptcle_contrib}(b).%\cite{hatch}
The linear dependence on hopping of the anomaly width is shown in the inset.

The physical mechanism behind the hard gap for this group of molecules is very simple:
When the LHO of one site is close in energy to the UHO on the neighboring site, a single electron placed in the system will largely sit on the lower energy site and the 1-particle ground state is essentially unchanged by $t$.  However, if a second electron is introduced it can spread itself over the two sites, lowering the 2-particle ground state energy by an amount linear in $t$.
LHO-UHO mixing without mixing of all four atomic orbitals is only possible when $U/t\gg1$, and in this sense this mechanism is specific to strongly correlated systems.

An analytic expression for the DOS contribution of these molecules can be developed to leading order in $t/U$.
The 16$\times$16 two-site Hamiltonian matrix breaks into diagonal blocks with fixed particle number.  Moreover, the 2-particle block further divides into a triplet block and a singlet block.
Because the states in the triplet block are not coupled by $t$ while those in the singlet block are, the 2-particle ground state is always a singlet and may be written as a linear combination of $c_{1\uparrow}^{\dag} c_{1\downarrow}^{\dag}|\rangle$, $c_{2\uparrow}^{\dag} c_{2\downarrow}^{\dag}|\rangle$, and ${1 \over \sqrt{2}}(c_{1\uparrow}^{\dag} c_{2\downarrow}^{\dag} - c_{1\downarrow}^{\dag} c_{1\uparrow}^{\dag})|\rangle$.
When $\epsilon_2 \sim \epsilon_1 + U$ and $U>>t$, the contribution of $c_{2\uparrow}^{\dag} c_{2\downarrow}^{\dag}|\rangle$ may be dropped to leading order in $t/U$.  The ground state energy is therefore given by the lowest eigenvalue of the matrix
\begin{eqnarray}
\left( \begin{array}{cc}
\epsilon_1 + \epsilon_2 & - \sqrt{2} t \\
-\sqrt{2} t & 2 \epsilon_1 + U \end{array} \right).
\end{eqnarray}
When $\epsilon_2>\mu>\epsilon_1+U$ ($|d_1\rangle$ in Fig.~\ref{fig:site_energies}(a)), the atomic ground state is $c_{1\uparrow}^{\dag} c_{1\downarrow}^{\dag}|\rangle$ 
and with hopping the ground state energy is
$E_g = 2\mu - U + 2x - y - \sqrt{y^2 + 2t^2}$,
where $x=(\epsilon_1 + \epsilon_2 + U - 2 \mu)/2$ 
and $y=(\epsilon_2 - \epsilon_1 - U)/2$.
%When $y=0$, the ground state energy is $\sqrt{2} t$ less than in the atomic limit.
When $y=0$, the ground state energy is lowered by $\sqrt{2} t$.

Equally important to the DOS, the energies of the 1- and 3-particle states are {\em not} shifted to leading order in $t/U$.  
%Keeping only terms of zeroth order in $t/U$ and dropping transitions with energies of order $U$, 
%Dropping terms of order $t/U$, 
At zeroth order in $t/U$,
the DOS near $\varepsilon_F$ of a single molecule in the $|d_1\rangle$ configuration is
\begin{eqnarray}
\rho(\omega) &=& 
{1 \over 2} F(y)
\biggl\{
\delta(\omega - x - \sqrt{y^2 + 2t^2})
\nonumber \\
& & \hskip 0.45 in
+ \delta(\omega - x + \sqrt{y^2 + 2t^2})
\biggr\}
\\
{\rm where} \ F(y) &=& {2 y^2 + 3 t^2 + 2 y \sqrt{y^2 + 2t^2}
\over 2 y^2 + 4 t^2 + 2 y \sqrt{y^2 + 2t^2}}.
\end{eqnarray}
The contribution of all $|d_1\rangle$ molecules is determined by integrating over 
$-{\Delta \over 2} < \epsilon_1 < \mu - U$ and $\mu < \epsilon_2 < {\Delta \over 2}$.
$F(y)$ depends weakly on $y$ and we set it to one, its limiting value for large $y$.
The two sites in a molecule are equivalent, so the contributions of $|d_1\rangle$ and $|d_2\rangle$ molecules are the same.
An analogous calculation may be done for the $|s\rangle$ molecules.
Combining these and normalizing by dividing by $\Delta^2$, we obtain the following disorder-averaged DOS due to the $|d\rangle$ and $|s\rangle$ molecules.
\begin{eqnarray}
{\bar \rho}(\omega)
&\approx& f(\omega,A_{d\alpha}) + f(\omega,A_{s\alpha})
\label{analyticeqn}
\\
{\rm where} \ f(\omega, A) 
&=& {(|\omega| + A)^2 - 2t^2 \over (|\omega|+A)} - \left|{|\omega|^2 - 2t^2 \over |\omega|} \right|
\nonumber
\end{eqnarray}
and where $\alpha={\rm sign}\ \omega$.
When
$U-\Delta/2< \mu \leq U/2$,  
$A_{d+} = (\Delta/2) - U + \mu$ and 
$A_{d-} = (\Delta/2) - \mu$ 
while $A_{s+} = A_{s-} = U/2$.
And when $(U-\Delta)/2 < \mu < U-\Delta/2$,
there are no molecules of type $|d\rangle$, and 
$A_{s+} = U/2$  and 
$A_{s-} = (\Delta/2) + \mu - (U/2)$.

The key point is that ${\bar \rho}(\omega)$ in Eqn.~(\ref{analyticeqn}) has a peak at $\sqrt{2} t$.
Fig.~\ref{fig:2ptcle_contrib}(c) shows this analytic expression as a function of $t$ at half filling for $\Delta=20$ and $U=12$.  
The linear dependence of the peak position on $t$ is clear.
Moreover, the fact that the position of the peak in the function $f(\omega,A)$ is independent of $A$ means that the width of the anomaly is independent of all remaining parameters---$U$, $\Delta$ and $\mu$---as seen in Fig.~\ref{fig:ensemble_dos} and in the lattice calculations.
At $\mu=U-\Delta/2$, two changes occur:  First, the atomic DOS is no longer flat at $\varepsilon_F$ but instead has a positive step, and, second, there are no more $|d\rangle$ molecules.  These effects are reflected in the $\mu=U/8$ curve in Fig.~\ref{fig:ensemble_dos}(d).  

This calculation retains only the kinetic energy savings of the 2-particle ground state, and the similarity of the results to those in Fig.~\ref{fig:2ptcle_contrib}(b) demonstrates that this is the physics underlying the anomaly in the molecular ensemble.  Moreover, the parallels between the behavior of the molecular ensemble and that of the lattice suggest that this may in fact be the physics underlying the energy scale $t$ of the lattice anomaly as well.

{\em Discussion}
We have focused here on the $|d\rangle$ and $|s\rangle$ molecules which make the dominant contribution to the anomaly for $\Delta>U$ near half filling. 
One might expect linear $t$ dependence from all molecules by the following logic:
For molecules with LHO-LHO alignment, the 1- and 3-particle atomic states are shifted in energy by an amount linear in $t$, while the 0-, 2- and 4-particle atomic states are unchanged to zeroth order in $t/U$.  
Only transitions between states which differ in particle number by one enter the single-particle DOS, so a linear dependence on $t$ would seem to arise in all cases.
In fact, when the contributions of all LHO-LHO aligned molecules are summed, there is a cancelation of all terms linear in $t$.
As a result, other than the $|s\rangle$ and $|d\rangle$ molecules the only molecules making a contribution to the ZBA with linear $t$ dependence are those with 1- and 3-particle atomic ground states and LHO-UHO alignment.  
In this case the addition of hopping can change not just the transition energies but also the number of allowed transitions, by causing a change in the number of particles in the ground state.
A more detailed discussion of the two-site ensemble will be provided elsewhere.\cite{Atkinson2010-2site}

The range of disorder and interaction strength to which this picture is applicable depends on coordination number, with higher disorder required in systems with higher coordination number.
For systems in which the disorder strength is much greater than the bandwidth and the interaction strength is much larger than the hopping, the framework presented here suggests the following:
First, in a strong magnetic field the alignment of electron spins would preclude the sharing of the second electron, reducing the depth of the anomaly but not its width with increasing field.
Second,
a number of experiments\cite{Sarma1998,Ray2006,Maiti2007} have observed hard gaps. 
In particular, the energy dependence of the anomaly seen at high disorder in Mn doped LaNiO\cite{Sarma1998} bears an intriguing resemblance to Fig.~\ref{fig:2ptcle_contrib}(b).  
Particular combinations of dopant and parent compound may result in more frequent occurance of the LHO-UHO alignment.
Finally, an anomaly of width $t$ is not expected for $U\gg\Delta$ where LHO-UHO alignment is not present.
In apparent contrast, Ref.~\onlinecite{Chiesa2008} states that a ZBA persists for $U\gg\Delta$.  
However, the width of the DOS suppression in the lattice when $U\gg\Delta$ is no longer $t$ and decreases with increasing disorder.\cite{private_communication}
The low energy physics of the anomaly and its behavior outside the parameter range to which the two-site ensemble is relevant--disorder less than or of order the bandwidth--remain important open questions.
We suggest that at low energies a mechanism analogous to the two-site picture may persist:  the mixing by hopping of many-body states, but in this case states which extend over multiple sites.  
%At large $U$, when the simple two-site mixing no longer contributes, these longer length scale effects continue to produce an anomaly, albeit with a reduced energy scale.

In summary, we have demonstrated how disorder can generate a feature in the DOS of strongly correlated systems with an energy scale given by the hopping, rather than the usual strong-coupling scale $J$.  
The origin of the effect is the kinetic energy savings allowed by spreading an electron between two neighboring sites.
The fact that the key mixing occurs between the LHO on one site and the UHO on the neighboring site gives this mechanism a character unique to strongly correlated systems.

We acknowledge support by NSERC, CFI and OIT.  
RW thanks Dr.~Simone Chiesa for useful discussions, and the Laboratoire de Physique des Solides of the University of Paris in Orsay and the Department of Physics at the University of California in Davis for hosting her.

%\bibliographystyle{apsrev}
%\bibliography{Disorder-Hubbard}

\begin{thebibliography}{23}
\bibitem[{sof()}]{softgap2}
\bibinfo{note}{Lanczos and quantum Monte Carlo methods do not have the energy
  resolution to resolve a soft gap of this size, and its existence in two
  dimensions is therefore not established.}
\bibitem[{pha()}]{phase_space_range}
\bibinfo{note}{Note that LHO-LHO alignment refers to the whole region
  $\epsilon_1-U/2< \epsilon_2 < \epsilon_1+U/2$ and not just the line
  $\epsilon_1=\epsilon_2$, and similarly for LHO-UHO alignment.}
\end{thebibliography}

\end{document}